\documentclass[useAMS, usenatbib]{mn2e}
\pdfoutput=1
\usepackage{graphicx}
\usepackage{longtable}
\usepackage{multicol}
\usepackage{lscape}
\usepackage{subfig}
\usepackage{natbib}
\usepackage{threeparttable}
\usepackage{gensymb}

\title[Intermediate Inclinations of Type 2 Coronal-Line Forest AGN]{Intermediate Inclinations of Type 2 Coronal-Line Forest AGN}
\author[M. Rose, M. Elvis, M. Crenshaw, A. Glidden]{Marvin Rose$^{1}$\thanks{E-mail:mrose@cfa.harvard.edu (MR)}, Martin Elvis$^{1}$, Michael Crenshaw$^{2}$, Ana Glidden$^{3}$ \\
$^{1}$Harvard Smithsonian Center for Astrophysics, 60 Garden St., Cambridge, MA 02138, USA\\
$^{2}$Department of Physics and Astronomy, Georgia State University, Astronomy Offices, 25 Park Place, Suite 600, Atlanta, GA 30303, USA\\
$^{3}$Department of Physics, Massachusetts Institute of Technology, 77 Massachusetts Avenue, Cambridge, MA 02139\\}
\begin{document}

\date{}

\pagerange{\pageref{firstpage}--\pageref{lastpage}} \pubyear{2011}

\maketitle

\label{firstpage}

\begin{abstract}

Coronal-Line Forest Active Galactic Nuclei (CLiF AGN) are remarkable in the sense that they have a rich spectrum of dozens of coronal emission lines (e.g. [FeVII], [FeX] and [NeV]) in their spectra. Rose, Elvis \& Tadhunter (2015) suggest that the inner obscuring torus wall is the most likely location of the coronal line region in CLiF AGN, and the unusual strength of the forbidden high ionization lines is due to a specific AGN-torus inclination angle. Here we test this suggestion using mid-IR colours (4.6$\mu$m-22$\mu$m) from the Wide-Field Infrared Survey Explorer (WISE) for the CLiF AGN. We use the Fischer et al. (2014) result that showed that as the AGN-torus inclination becomes more face on, the Spitzer 5.5$\mu$m to 30$\mu$m colours become bluer. We show that the [W2-W4] colours for the CLiF AGN ($\langle$[W2-W4]$\rangle$ = 5.92$\pm$0.12) are intermediate between SDSS type 1 ($\langle$[W2-W4]$\rangle$ = 5.22$\pm$0.01) and type 2 AGN ($\langle$[W2-W4]$\rangle$ = 6.35$\pm$0.03). This implies that the AGN-torus inclinations for the CLiF AGN are indeed intermediate, supporting the work of Rose, Elvis \& Tadhunter (2015). The confirmed relation between CLiF AGN and their viewing angle shows that CLiF AGN may be useful for our understanding of AGN unification. 
 
\end{abstract}

\begin{keywords}
galaxies:active --- galaxies:Seyfert --- quasars:emission lines --- quasars:general 
\end{keywords}

\section{Introduction}

The nature and origin of the dusty obscuring torus believed to surround Active Galactic Nuclei (AGNs) - whether it is inflowing, outflowing, and its relation to star formation - are important for our understanding of the evolution of AGN, and how they affect their environment (`feedback'). However, the nature of the torus is problematic because direct observations are rare.

Recently, Rose, Elvis \& Tadhunter (2015; RET15 hereinafter) introduced the Coronal Line Forest Active Galactic Nuclei (CLiF AGN). The optical spectra of CLiF AGN are remarkable. As well as typical low ionization AGN emission lines (FLILs; e.g. [OIII], [OII], [NII], [SII]) and Balmer recombination lines (e.g. H$\alpha$, H$\beta$), their spectra include dozens of strong forbidden high ionization emission lines (FHILs; e.g. [FeVII], [FeX], [NeV] and [ArV]). RET15 define the boundary between low and high ionization to be the ionization potential (I$_{P}$) of the HeII edge at 54.4 eV. Six out of seven of the CLiF AGN in RET15 lacked broad emission lines (FWHM H$\beta$ $>$ 2000 km s$^{-1}$) , i.e. they are type 2 AGN, and so are likely observed with our line of sight at a relatively large inclination angle to the torus axis. The far wall of the torus is a plausible source for the FHILs which will be strongest if we have a maximal view of the inner wall (Nagao et al. 2001; Rose et al. 2011). The Coronal Line Region (CLR) lies at a distance comparable to the hot dust sublimation radius (RET15). RET15 use photoionization modelling results (CLOUDY; Ferland 1998) to calculate the CLR radial distances for the CLiF AGN and find that they are comparable to the dust grain sublimation distances, thus supporting this view. 

If this picture proves correct, then the geometry of these AGNs is unusually well-constrained, making them laboratories to test AGN torus models.

Fischer et al. (2014) have recently shown that the Spitzer mid-IR (F$_{5.5}$/F$_{30}$) colours change smoothly with the AGN inclination angle derived from kinematic models of AGN bi-cones. Crenshaw et al. (in prep.) confirmed this trend using Wide-Field Infrared Survey Explorer (WISE; \citealt{wright10}) mid-IR colours [W2-W4] (4.6$\mu$m-22$\mu$m). This result allows us to use [W2-W4] as a proxy for inclination and so test whether CLiF AGN have the special orientation suggested by RET15.

\begin{center}
\begin{table*}
\centering
\caption{Redshifts, WISE photometry results (mag; \citealt{wright10}) and uncertainties for the type 2 CLiF AGN. Note that there are no warning flags for the photometry of this sample, and all the quality flags indicate that the sources were detected with S/N $>$ 10.}
\begin{tabular}{lcccccc}
\hline
Name	&	z	&	W1	&  W2	& W3	& W4 & [W2-W4] \\
\hline
ESO 138 G1	&	0.0091	&	8.82$\pm$0.02	&	7.58$\pm$0.02	&	4.10$\pm$0.02	&	1.64$\pm$0.01	&	5.94$\pm$0.03		\\
Tololo 0109-383	&	0.0118	&	8.09$\pm$0.02	&	6.86$\pm$0.02	&	4.03$\pm$0.02	&	1.91$\pm$0.01	&	4.95$\pm$0.03		\\
J1142+14	&	0.0207	&	11.29$\pm$0.02	&	10.29$\pm$0.02	&	6.74$\pm$0.02	&	4.07$\pm$0.03	&	6.22$\pm$0.05		\\
Mrk 1388	&	0.0216	&	10.83$\pm$0.02	&	9.70$\pm$0.02	&	6.27$\pm$0.02	&	3.90$\pm$0.02	&	5.80$\pm$0.04		\\
J1440+02	&	0.0299	&	13.39$\pm$0.02	&	12.26$\pm$0.02	&	7.77$\pm$0.02	&	4.54$\pm$0.03	&	7.72$\pm$0.05		\\
J1109+28	&	0.0329	&	11.54$\pm$0.02	&	10.38$\pm$0.02	&	7.01$\pm$0.02	&	4.49$\pm$0.03	&	5.89$\pm$0.05		\\
J1241+44	&	0.0422	&	13.85$\pm$0.03	&	13.25$\pm$0.03	&	9.61$\pm$0.04	&	7.21$\pm$0.10	&	6.04$\pm$0.13		\\
J1056+44	&	0.0525	&	12.91$\pm$0.02	&	11.17$\pm$0.02	&	8.17$\pm$0.02	&	5.84$\pm$0.04	&	5.33$\pm$0.06		\\
J1407+10	&	0.0601	&	12.00$\pm$0.02	&	10.81$\pm$0.02	&	7.34$\pm$0.02	&	5.02$\pm$0.03	&	5.79$\pm$0.05		\\
J1609+04	&	0.0639	&	11.42$\pm$0.02	&	10.00$\pm$0.02	&	6.18$\pm$0.02	&	3.68$\pm$0.02	&	6.32$\pm$0.04		\\
J1316+44	&	0.0906	&	10.01$\pm$0.02	&	8.86$\pm$0.02	&	5.66$\pm$0.02	&	3.24$\pm$0.02	&	5.62$\pm$0.04		\\
J0858+31	&	0.1389	&	12.47$\pm$0.02	&	11.15$\pm$0.02	&	7.65$\pm$0.02	&	4.98$\pm$0.04	&	6.17$\pm$0.06		\\
Q1131+16	&	0.1732	&	12.43$\pm$0.02	&	11.12$\pm$0.02	&	7.94$\pm$0.02	&	5.78$\pm$0.05	&	5.34$\pm$0.07		\\
J1641+43	&	0.2214	&	12.92$\pm$0.02	&	11.66$\pm$0.02	&	8.29$\pm$0.02	&	5.59$\pm$0.04	&	6.07$\pm$0.06		\\
\hline
\end{tabular}
\label{tab:gen}
\end{table*}
\end{center}

\section{CLiF AGN Sample}\label{sect:samp}

In addition to the 6 type 2 CLiF AGN from RET15 we use another 8 found by Rose et al. (2015, in prep.). RET15 showed that the CLiF AGN lie in the sparsely populated cleft between the two branches of the BPT diagnostic diagram that dictate whether the emission region of an object is photoionized by an AGN, or by a starburst (\citealt{bpt}; \citealt{veilleux87}; \citealt{kewley06}), due most likely to collisional excitation of H$\alpha$ in a high density (n$_e$ $>$ 10$^{6.0}$ cm$^{-3}$), partially ionized zone. Rose et al. (2015, in prep.) use this region to select additional CLiF AGN from the SDSS galaxy and quasar catalogs. This increases the known sample of CLiF AGN (type 1 and 2) to 99 objects. 

Rose et al. (2015, in prep.) also find 84 type 1 CLiF AGN (FWHM H$\beta$ $>$ 2000 km s$^{-1}$). The nature of these objects are distinct from the type 2 CLiF AGN and will be discussed in Rose et al. (2015, in prep.) and, briefly, in the discussion section of this paper. 

The redshift range seen for the entire sample of CLiF AGN is 0.02 $<$ z $<$ 0.30, as H$\alpha$ must lie within the SDSS DR10 spectral range.

In Table \ref{tab:gen} we give the redshifts, WISE photometry and [W2-W4] colours for the 14 type 2 CLiF AGN. 

\section{Results}\label{sect:wise}

\subsection{WISE colours as an indicator of AGN inclination}

The AGN-torus inclinations of nearby (z $<$ 0.06) AGN can be determined by mapping the kinematics of their narrow-line regions (NLRs), using the [OIII]$\lambda$5007 emission line with high spatial resolution long-slit  
Hubble Space Telescope-Space Telescope Imaging Spectrograph or Integral Field Unit spectra. NLR kinematics are found to be dominated by radial outflows which can be well fitted with simple biconical outflow models. 

If the AGN-torus and NLR bi-cone axes are co-aligned then NLR inclinations would be expected to correlate with mid-IR colours if the torus is optically thick at these wavelengths. In general this appears to be a good assumption (\citealt{ck00}; \citealt{crenshaw00}; \citealt{fischer13}).

In Figure \ref{fig:24} we show a plot of NLR bi-cone inclination and WISE mid-IR colours W2 (4.6$\mu$m) - W4 (22$\mu$m) which appear correlated (R=0.72, p $<$ 0.001) although with scatter. The data for type 1 (blue circles) and type 2 AGN (red triangles) are from Fischer et al. (2014; Table \ref{tab:c15}). There are no CLiF AGN in this sample. The W2 and W4 bands are chosen because these bands do not coincide with any host galaxy features which could affect their colours, e.g. PAH features coincide with the W3 (12$\mu$m) band. [W2-W4] would then become bluer with more face-on inclinations because the hottest dust (T $>$ 500K) from the torus, whose emission peaks at shorter (i.e. W2) wavelengths, becomes less obscured by the cooler dust dominating in the W4 band (T $\sim$ 150K).

Overall, type 2 AGN have redder [W2-W4] colours when compared to type 1 AGN. However, this simple picture could be complicated by sources of scatter, e.g. different opening angles, torus clumpiness or line-of-sight differences. For example, the object with the bluest W2-W4 colour is a type 2 AGN. This scatter is too wide (see $\sigma$$_{[W2-W4]}$ in Table \ref{tab:w24med}) to derive an angle from the [W2-W4] colour for specific objects. 

\subsection{WISE colours of Type 2 CLiF AGN}

RET15 suggested that CLiF AGN are observed with our line of sight at a large angle at which much of the inner wall of the torus can be seen, but the continuum and BLR remain hidden. This suggests that they are viewed at an intermediate inclination when compared to type 1 and 2 AGN, a hypothesis which can be tested using their WISE [W2-W4] colours as a proxy for AGN inclination.

In this section we will compare the [W2-W4] colours for three samples: a sample of SDSS type 2 AGN (\citealt{reyes08}; 420 objects), SDSS type 1 AGN sample (\citealt{shen11}; 815 objects), and type 2 CLiF AGN (RET15; Rose et al. 2015, in prep).  

Note that we only used objects which have full WISE (W1-W4) photometry and that there are no warning flags for the photometry, and all the quality flags indicate that the sources were detected with S/N $>$ 10. In each case we limited the redshifts of the samples to z $<$ 0.3 so that they are consistent with the CLiF AGN sample. Note there is no overlap between these three samples. 

We use these samples for comparison with the type 2 CLiF AGN (rather than the objects shown in Figure 1) because, like the type 2 CLiF AGN, they are selected from the SDSS using their emission line properties. The type 2 AGN and type 2 CLiF AGN are selected from both the main SDSS galaxy and quasar catalogs. 75\% of the SDSS type 2 AGN with z $<$ 0.30 are selected from the SDSS galaxy catalog \citep{reyes08}, compared to 71\% of the type 2 CLiF AGN. Therefore differences in their mid-IR colours should not be the result of their selection.

The type 1 AGN have only been selected from the SDSS quasar catalog. The quasar catalog uses optical colour cuts which ensures the bluest objects are selected \citep{richards02}. Given the nature of type 1 AGN (amongst the bluest objects in the Universe) this is entirely appropriate. While bluer optical colour criteria may coincide with bluer mid-IR colours, we cannot rule out that these colours are the result of orientation effects, rather than a selection bias.   

\begin{figure}
\centering
\includegraphics[scale=0.38, trim=0.1cm 4.6cm 0.1cm 3.6cm]{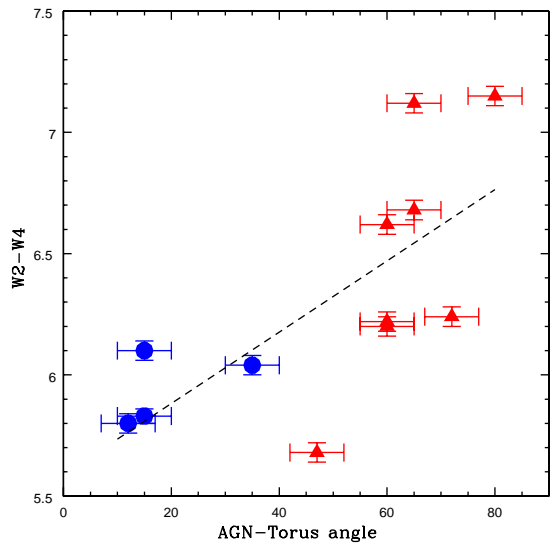}
\caption{[W2-W4] colours versus AGN inclination for the \citet{fischer14} sample of AGN. Type 1 AGN are represented with blue circles and the type 2 AGN are represented with red triangles. The dashed line presents the best fit line and has parameters y=0.015x+5.59.}
\label{fig:24}
\end{figure}

\begin{figure}
\centering
\includegraphics[scale=0.38, trim=0.1cm 4.6cm 0.1cm 3.6cm]{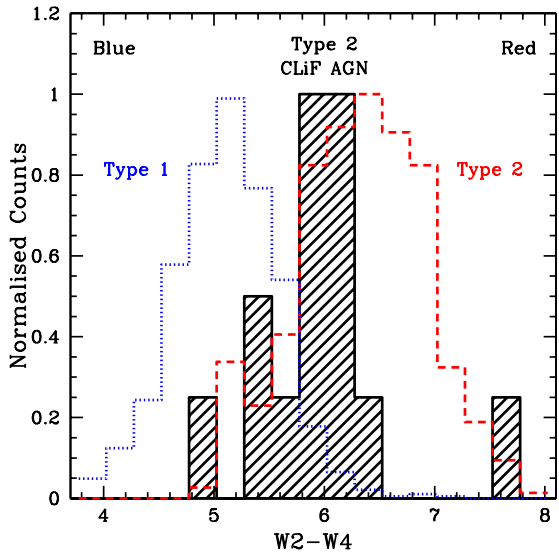}
\caption{Normalised histograms showing the [W2-W4] colours for the redshift limited (z $<$ 0.30) \citet{reyes08} sample of type 2 AGN (red dashed line), redshift limited (z $<$ 0.30) \citet{shen11} type 1 AGN sample (blue dotted line) and type 2 CLiF AGN (black solid line). The bin size is 0.25 magnitudes.}
\label{fig:w24}
\end{figure}

In Figure \ref{fig:w24} we plot normalised histograms for the WISE [W2-W4] colours of the three samples: SDSS type 2 AGN (red dashed line), SDSS type 1 AGN (blue dotted line), and type 2 CLiF AGN (black solid line). The [W2-W4] colours of the type 1 AGN are clearly bluer when compared to the type 2 AGN. This is supported by the median [W2-W4] colours for each sample (Table \ref{tab:c15}). The implied AGN-torus inclinations in the correlation shown in Figure \ref{fig:24}, agree with these results: type 1 AGN clearly have face-on inclinations and type 2 clearly favour edge-on inclinations. The median [W2-W4] colours for the SDSS type 1 AGN are also bluer than the type 1 AGN studied in \citet{fischer14}. But, as discussed above, this could be due to a selection effect. 

Strikingly, the median [W2-W4] colour for the type 2 CLiF AGN is intermediate between the \citet{reyes08} sample of type 2 AGN and the type 1 AGN sample (Figure \ref{fig:w24}). Indeed type 2 CLiF AGN seem to fill the dip between the type 1 and type 2 samples. To quantify the significance of these differences we performed KS tests comparing [W2-W4] for the 3 samples (Table \ref{tab:ksw24}). The results of these tests strongly support the differences seen in the median colours of the AGN types. No sample is consistent with the other at more than 0.4\%. However, the type 2 CLiF AGN sample size is small (14) and normality tests (e.g. KS tests) become less reliable for small samples (n $<$ 20; \citealt{saculinggan13}). 

We note that the median [W2-W4] colour of the type 2 CLiF AGN is comparable to that of the Fischer et al. (2014) sample of type 1 AGN. However, the Fischer et al. (2014) sample of type 1 AGN is much smaller (4 objects) when compared to the SDSS-selected type 1 AGN (815 objects). In addition, the SDSS-selected type 1 AGN are a more appropriate comparison sample for the CLiF AGN because both samples have been selected from the SDSS, and have comparable redshift ranges.

There is significant scatter in the [W2-W4] colours for all the samples ($\sigma$$_{[W2-W4]}$ in Table \ref{tab:w24med}) resulting in an overlap in colour among the various types. However as explained in $\S$3.1, the scatter could be due to a variety of causes. For example different opening angles, torus clumpiness or line-of-sight differences. 

The overall [W2-W4] colour is clearly bluer as the AGN type changes from type 2, through CLiF type 2 to the type 1 AGN.

\begin{center}
\begin{table}
\centering
\caption{AGN type, inclination ($\theta$) and [W2-W4] colours for the AGN studied in \citep{fischer14}.}
\begin{tabular}{lccccccc}
\hline
Name	&	AGN Type	&	$\theta$$\degree$	&  [W2-W4] \\
\hline
NGC	4507	&	2	&	47$\pm$5	&	5.68$\pm$0.04		\\
NGC	4051	&	1	&	12$\pm$5	&	5.80$\pm$0.04		\\
NGC	3783	&	1	&	15$\pm$5	&	5.83$\pm$0.03		\\
Mrk	279	&	1	&	35$\pm$5	&	6.04$\pm$0.04		\\
NGC	3227	&	1	&	15$\pm$5	&	6.10$\pm$0.04		\\
Mrk	573	&	2	&	60$\pm$5	&	6.20$\pm$0.04		\\
NGC	7674	&	2	&	60$\pm$5	&	6.22$\pm$0.04		\\
NGC	1667	&	2	&	72$\pm$5	&	6.24$\pm$0.04		\\
Mrk	78	&	2	&	60$\pm$5	&	6.62$\pm$0.04		\\
Mrk	34	&	2	&	65$\pm$5	&	6.68$\pm$0.04		\\
NGC	5643	&	2	&	65$\pm$5	&	7.12$\pm$0.04		\\
Mrk	1066	&	2	&	80$\pm$5	&	7.15$\pm$0.04		\\
\hline
\end{tabular}
\label{tab:c15}
\end{table}
\end{center} 

\begin{center}
\begin{table}
\centering
\caption{Median [W2-W4] colours and standard deviations on the [W2-W4] colours for the redshift limited (z $<$ 0.30) \citet{reyes08} sample of type 2 AGN (T2), redshift limited (z $<$ 0.30) \citet{shen11} type 1 AGN sample (T1), type 2 CLiF AGN (T2$_C$), type 1 CLiF AGN (T1$_C$), type 1 AGN of \citet{fischer14} (T1$_F$) and type 2 AGN of \citet{fischer14} (T2$_F$).}
\begin{tabular}{lcc}
\hline
Type	&	[W2-W4] & $\sigma$$_{[W2-W4]}$	\\
\hline			
T1	&	5.22$\pm$0.01	&	0.47 \\
T1$_C$	&	5.28$\pm$0.04	&	0.36 \\
T2$_C$	&	5.92$\pm$0.11	&	0.64 \\
T2	&	6.35$\pm$0.03	&	0.58 \\
\hline
T1$_F$  &       5.93$\pm$0.07   &       0.15 \\
T2$_F$  &       6.43$\pm$0.15   &       0.50 \\
\hline																			
\end{tabular}
\label{tab:w24med}
\end{table}
\end{center} 

\begin{center}
\begin{table}
\centering
\caption{KS test results (P-value) comparing the [W2-W4] distribution of the redshift limited (z $<$ 0.30) \citet{reyes08} sample of type 2 AGN (T2), redshift limited (z $<$ 0.30) \citet{shen11} type 1 AGN sample (T1), type 2 CLiF AGN (T2$_C$) and type 1 CLiF AGN (T1$_C$).}
\begin{tabular}{lcccc}
\hline
	&	T1	&	T1$_C$	&	T2$_C$	&	T2	\\
\hline
T1	&	-	&	0.22	&	$<$0.0001	&	$<$0.0001	\\
T1$_C$	&		&	-	&	$<$0.0001	&	$<$0.0001	\\
T2$_C$	&		&		&	-	&	0.004	\\
\hline
\end{tabular}
\label{tab:ksw24}
\end{table}
\end{center}

\section{Discussion}\label{sect:disc}

\begin{figure}
\centering
\includegraphics[scale=0.38, trim=0.1cm 4.6cm 0.1cm 3.6cm]{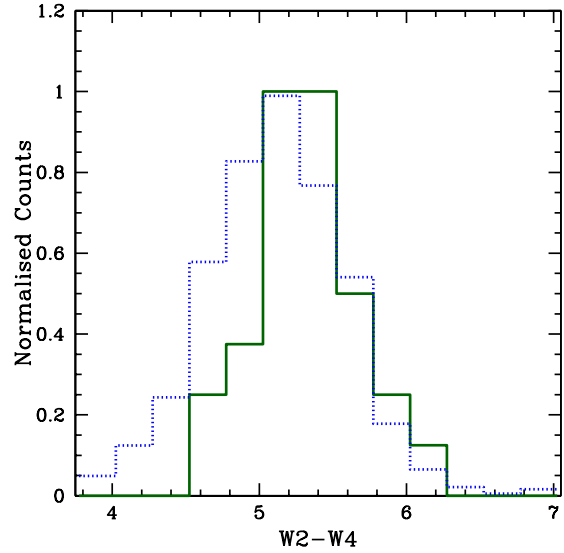}
\caption{Normalised histograms showing the [W2-W4] colours for the redshift limited (z $<$ 0.30) \citet{shen11} type 1 AGN sample (blue dotted line) and type 1 CLiF AGN (green solid line). The bin size is 0.25 magnitudes.}
\label{fig:w24b}
\end{figure}

Assuming that the overall differences in [W2-W4] for the AGN studied in Figure \ref{fig:w24} are dominated by the inclination angle (Figure \ref{fig:24}), and not by selection effects, the [W2-W4] colours imply that type 2 CLiF AGN lie at an intermediate inclination angle when compared to the samples of SDSS type 1 and type 2 AGN. This relatively large angle to the torus axis may well provide us with a maximal view of the far wall of the torus, while the AGN continuum is still obscured by the near side of the torus, as suggested by RET15. Type 2 CLiF AGN may be useful for our understanding of AGN unification. 

It is instructive to  see where the new type 1 CLiF AGN line in [W2-W4] colour space. In Figure \ref{fig:w24b} we plot normalised histograms showing the [W2-W4] colours for the SDSS type 1 AGN sample (\citealt{shen11}; blue dotted line) and type 1 CLiF AGN (Rose et al. 2015, in prep; green solid line). The [W2-W4] colours for the typical type 1 AGN and type 1 CLiF AGN are indistinguishable (P-value = 0.22; Table \ref{tab:ksw24}). This is supported by the median values presented in Table \ref{tab:w24med} as the median [W2-W4] (and therefore AGN-torus inclinations) agree within two error bars. A KS test comparing [W2-W4] for these samples (Table \ref{tab:ksw24}) supports the view that there is no difference in their [W2-W4] colours. All the type 1 CLiF AGN were selected from the quasar catalog, therefore these comparisons are not biased by selection effects. In addition, there is no overlap between the samples. The lack of significant differences between the [W2-W4] colours of the type 1 CLiF AGN and the SDSS type 1 AGN implies that the type 1 CLiF AGN have no special inclination when compared to typical type 1 AGN. Type 1 CLiF AGN are therefore distinct from type 2 CLiF AGN. Their strong CLR must then be due to another distinct mechanism. 

We will study both these populations of CLiF AGN in further papers in the following ways.

\begin{enumerate}

\item[1.] Model the inner torus wall geometry proposed to produce strong FHILs (Glidden et al. 2015, in prep.).

\item[2.] Determine the physical conditions (ionization parameter and electron density) of the CLR, and estimate the radial distances to this region (Rose et al. 2015, in prep.).

\end{enumerate}

In the longer term it would clearly be extremely valuable to obtain high spatial resolution spectroscopy for the CLiF AGN, so that the AGN-torus inclination of type 2 CLiF AGN can be estimated using the NLR bicones \citep{fischer14}.

\section*{Acknowledgments}

This work was supported in part by NASA grant NNX13AE88G. We thank the referee for useful comments and suggestions which improved this work substantially. The authors acknowledge the data analysis facilities provided by the Starlink Project, which is run by CCLRC on behalf of PPARC. This publication makes use of data products from NEOWISE, which is a project of the Jet Propulsion Laboratory/California Institute of Technology, funded by the Planetary Science Division of the National Aeronautics and Space Administration. This research has made use of the NASA/IPAC Extragalactic Database (NED) which is operated by the Jet Propulsion Laboratory, California Institute of Technology, under contract with the National Aeronautics and Space Administration. Funding for SDSS-III has been provided by the Alfred P. Sloan Foundation, the Participating Institutions, the National Science Foundation, and the U.S. Department of Energy Office of Science.

\end{document}